\documentclass[aps,twocolumn,preprintnumbers,showpacs,showkeys]{revtex4}
\usepackage{epsfig}
\usepackage{amssymb,amsmath,amsfonts,amsthm,graphicx}
\newcommand{\beq}{\begin{equation}}
\newcommand{\eeq}{\end{equation}}
\newcommand{\bea}{\begin{array}}
\newcommand{\eea}{\end{array}}
\newcommand{\beqa}{\begin{eqnarray}}
\newcommand{\eeqa}{\end{eqnarray}}

\def\beqa{\begin{eqnarray}}
\def\eeqa{\end{eqnarray}}

\begin{document}

\title{Density and correlations of topological objects \\near the transition temperature in lattice gluodynamics}

\author{V.~G.~Bornyakov}
\affiliation{Institute for High Energy Physics NRC ``Kurchatov Institute'',
142281 Protvino, Russia, \\
Institute of Theoretical and Experimental Physics, 117259 Moscow, 
Russia \\
School of Biomedicine, Far East Federal University, 690950 Vladivostok, 
Russia}

\author{E.-M. Ilgenfritz}
\affiliation{Joint Institute for Nuclear Research, BLTP, 141980 Dubna, Russia}

\author{B.~V.~Martemyanov}
\affiliation{Institute of Theoretical and Experimental Physics, 117259 Moscow, 
Russia \\
National Research Nuclear University MEPhI, 115409, Moscow, Russia \\
Moscow Institute of Physics and Technology, 141700, Dolgoprudny, Moscow Region,
Russia}
\date{\today}

\begin{abstract}
Topological lumps are known to be present in gluonic fields of $SU(3)$ gluodynamics.
Near the transition temperature they were classified either as constituents of nondissociated
(anti)calorons, or as constituents of (anti)dyon pairs, or as isolated (anti)dyons. 
In this paper we study the density and correlation functions of these objects at temperature 
$T/T_c=0.96$.
\end{abstract}

\keywords{Lattice gauge theory, overlap Dirac operator, caloron, dyon}

\pacs{11.15.Ha, 12.38.Gc, 12.38.Aw}

\maketitle

\section{Introduction}
\label{sec:introduction}

In Ref.~\cite{Ilgenfritz:2013oda} we have investigated topological objects 
formed in gluonic fields of $SU(3)$ gluodynamics near the transition temperature.
It was done with the help of  low lying  modes of the overlap Dirac operator.
These modes allow to construct the topological charge density corresponding to the three 
constituents (dyons) of a caloron with nontrivial holonomy~\cite{Kraan:1998pm,Kraan:1998sn,Lee:1998bb} 
when three types of fermionic temporal boundary conditions are used. 
 
The
dyons are playing the decisive role in a recent model of the QCD vacuum proposed by
Shuryak and collaborators 
~\cite{Shuryak:2011aa,Faccioli:2013ja,Larsen:2015vaa,Larsen:2014yya, Larsen:2017sqm}.
The density of dyons and their interaction determined in lattice simulations are important 
inputs for this model.
This motivated us to return to investigation of $SU(3)$ pure gluodynamics. We will present 
results obtained just below the confinement-deconfinement phase transition at $T/T_c=0.96$. 
In particular, we present a numerical value for the dyon number density and compare it with the model 
prediction~\cite{Larsen:2015vaa}. Results for various dyon correlation functions are presented 
for the first time.  

We study the density and interaction of dyons using the same method to construct the 
topological charge density as in Ref.~\cite{Ilgenfritz:2013oda}.
The new element introduced into this method is the new criterion to determine the number of 
low lying modes of the overlap Dirac operator that should be used for the construction of
the UV-filtered fermionic topological charge density.

In Section~\ref{sec:thermal-ensembles} the details of the lattice ensemble created at a
temperature near the deconfining transition are described. 
In Section~\ref{sec:clusters} we sketch the fermionic construction of the topological charge 
densities as applied for three types of fermionic temporal boundary conditions. 
A cluster analysis of the resulting topological charge densities provides us with a possiblity 
to localize dyons of different types. The results for dyon densities and dyon correlation functions 
are presented and discussed. Finally, we present our conclusions in Section~\ref{sec:conclusions}.

\section{Setup of the investigation}
\label{sec:thermal-ensembles}

The $SU(3)$ gauge field configurations for this investigation have been generated on a lattice 
of size $24^3 \times 6$ by sampling the pure $SU(3)$ gauge theory using the 
L\"uscher-Weisz action ~\cite{Luscher:1984xn}. 

In addition to the plaquette term (pl), the L\"uscher-Weisz action includes 
a sum over all $2 \! \times \! 1$  rectangles (rt) and a sum over all 
parallelograms (pg), i.e.~all possible closed loops of length 6 along the 
edges of all 3-cubes
\beqa
S[U]  = & \beta & \left(\sum_{pl} \frac{1}{3}  \mbox{Re~Tr}  [ 1 - U_{pl} ]\right.
\nonumber 
\\ 
& + &  c_1 \sum_{rt} \frac{1}{3}  \mbox{Re~Tr}  [ 1 - U_{rt} ] 
\\
& + & 
\left. c_2 \sum_{pg} \frac{1}{3}  \mbox{Re~Tr}  [ 1 - U_{pg} ] \right)\,,
\nonumber 
\label{sgauge}
\eeqa
where $\beta$ is the principal inverse coupling parameter, while the coefficients
$c_1$ and $c_2$ are computed using results of one-loop perturbation theory and 
tadpole improvement~\cite{Luscher:1985zq,Snippe:1997ru,Lepage:1992xa}:
\beq
c_1 =  -  \frac{1}{ 20  u_0^2}  
[ 1 + 0.4805  \alpha ]\,,~~
c_2 =  -  \frac{1}{u_0^2}  0.03325  \alpha \,.
\eeq
For a given $\beta$, the tadpole factor $u_0$  and the lattice coupling constant
$\alpha$ are self-consistently determined in terms of the average plaquette 
\beq
u_0  =  \Big( \langle \frac{1}{3} \mbox{Re~Tr}~U_{pl} \rangle 
\Big)^{1/4}\,, \quad \alpha  =  -
\frac{ \ln \Big( \langle \frac{1}{3} \mbox{Re~Tr}~U_{pl} \rangle 
\Big)}{3.06839} 
\eeq 
in the course of a series of iterations.

The ensemble of 100 configurations for the overlap analysis has been generated
at $\beta = 8.20$. According to previous work~\cite{Gattringer:2002mr}, this ensemble 
corresponds to a temperature of
$T = 287$ MeV
\footnote{The scale was fixed by setting the Sommer parameter to $r_0=0.5$~fm.}
close to the phase transition 
temperature $T_c = 300$ MeV \cite{Gattringer:2002dv}. 

\section{Topological clustering}
\label{sec:clusters}

We have analyzed the configurations of the ensemble by identifying and investigating $N \le 30$ 
zero and near-zero eigenmodes
of the overlap Dirac operator. The spectral analysis has been performed for three 
types of temporal boundary conditions (b.c.) 
applied to the fermion field $\psi$: 
\beq 
\psi (1/T) = \exp(i\phi)\psi(0)
\label{eq:bc1}
\eeq 
with
\beq
\phi = \left\{
\begin{array}{ll}
 \phi_1  \equiv -\pi/3\,, \\
 \phi_2  \equiv +\pi/3\,, \\
 \phi_3  \equiv ~~~~\pi\,. \\
\end{array} 
\right.
\label{eq:bc2}
\eeq
For these three types of b.c.'s  the fermionic zero mode is maximally localized at 
one of the  three constituent dyons in the case of a single-caloron solution with 
maximally nontrivial holonomy.
For each of these b.c.'s we have determined the topological index and 
have checked that it was independent of the choice of $\phi$.
The obtained spectra are also independent of b.c.'s and have a nonzero spectral 
density around zero value (signalling spontaneous violation of chiral symmetry).

In order to proceed further, we have reconstructed from the zero and 
non-zero modes the profiles of the UV-filtered topological charge density
corresponding to the chosen fermionic boundary condition
according to its spectral representation (for details 
see \cite{Hasenfratz:1998ri,Ilgenfritz:2007xu})
\beq
q_{i,N}(x) = - \sum_{j=1}^N 
\left( 1 - \frac{\lambda_{i,j}}{2} \right) \psi^{\dagger}_{i,j}(x)\gamma_5 
\psi_{i,j}(x)\,,
\label{eq:truncated_density}
\eeq
where $j$ enumerates the eigenvalues $\lambda_{i,j}$ equal and closest to zero. 
These precise eigenvalues $\lambda_{i,j}$, as well as the corresponding modes
$\psi_{i,j}(x)$, are 
characterized by the $i$-th boundary condition.
Correspondingly, the UV-filtered topological density $q_{i,N}(x)$ depends on 
the boundary condition, too.

We have applied a cluster analysis with a variable lower cut-off $|q_{i,N}(x)| > q_{\rm cut}>0$
to these density functions. The cut-off $q_{\rm cut}$ has been chosen such as to resolve 
a maximal number of internally connected (while mutually separated clusters). It has been 
independently adapted for each configuration. The purpose of the cluster analysis was to discover 
extended objects that we are going to consider as dyon candidates  (of the respective type).

We have found the following average numbers of clusters per configuration (comprising all 100 
configurations) corresponding to the three boundary conditions $i=1,2,3$:
$$ N_{1,30} = 27.0 (4), \quad N_{2,30} = 27.0 (4), \quad N_{3,30} = 27.0 (4)\,$$
which completely agree
within errors (the latter given in parentheses). 
Therefore in the confining (center symmetric) phase the abundance of all three types of clusters 
is equal, and the clusters can be interpreted as
dyons with maximally nontrivial holonomy.
For further comparison we made the same calculations with $N=10$ or $N=20$ low lying modes 
taken into account for the reconstruction of the topological density

$$ N_{1,10} = 14.6 (3), \quad N_{2,10} = 14.2 (3), \quad N_{3,10} = 14.1 (3)\,$$
$$ N_{1,20} = 21.0 (3), \quad N_{2,20} = 21.2 (4), \quad N_{3,20} = 21.0 (4)\,$$

While the mutual equality of multiplicity remains, the total number of clusters
changes monotonically with the number of analyzing modes.

Taking the lattice spacing value $a=0.115$~fm from Ref.~\cite{Gattringer:2002mr} 
we obtain the physical 
three-dimensional
dyon cluster density 
$\rho = (N_{1,30} + N_{2,30} + N_{3,30})/(24a)^3 = 3.9(6)$~fm$^{-3}$.

Next we have checked whether clusters of different types are correlated
among themselves.
We found (in average per configuration) 
$$ N_{d,30} = 40(1),  ~2N_{dd,30} = 24.3(7), ~3N_{ddd,30} = 16.7(7),$$
where $N_{d,30}$ is the number of isolated clusters, $N_{dd,30}$ is
the  number of pairs of connected clusters and $N_{ddd ,30}$ is the number of triplets
of connected clusters (full calorons). The clusters of different type were counted as connected 
in pairs or triplets if the distance 
between them
was less than two lattice spacings. 
Interpreted in terms of calorons of non-trivial holonomy this means
that we see full caloron-like clusters consisting of three constituents
on one hand and also completely dissolved caloron constituents on the other
hand.

Again for further comparison we made the same calculations with $N=10,20$ low lying modes
taken into account

$$ N_{d,10} = 22(1),~  2N_{dd,10} = 11.6(4), ~ 3N_{ddd,10} = 9.1(5)\,$$
$$ N_{d,20} = 32.4(9),~  2N_{dd,20} = 17.4(6), ~ 3N_{ddd,20} = 13.3(7)\,$$

We computed the topological susceptibility $\chi=<Q^2>/V_4$ 
(with $Q$ being the topological charge of configuration and $V_4$ being its
$4d$ volume)
and found  $\chi=(187\pm 6~\mathrm{MeV} )^4$. This result is in very good agreement with that of
Ref.~\cite{Gattringer:2002mr} obtained at same temperature and for same action as in our work.
Our modelling of the
topological susceptibility by ensemble of dyons, dyon pairs and full calorons
gives rise to 
$$\chi_{\rm model} = Q_d^2 n_{d,N} + Q_{dd}^2 n_{dd,N} + Q_{ddd}^2 n_{ddd,N}$$
equal to $(169\pm 2~\mathrm{MeV} )^4$,  
$(187\pm 2~\mathrm{MeV} )^4$ and $(201\pm 2~\mathrm{MeV} )^4$ for 
$N=10, 20, 30,$  respectively.
Here
$n_{d,N}=N_{d,N}/V_4,~n_{dd,N}=N_{dd,N}/V_4,~n_{ddd,N}=N_{ddd,N}/V_4~$ are 
the densities of isolated dyons, 
dyons pairs and full calorons respectively and $Q_d=\pm \frac{1}{3}$, 
$Q_{dd}=\pm \frac{2}{3}$,
$Q_{ddd}=\pm 1$ are their modelled topological charges.  
As one can see the best agreement of our modelling with the result of Ref.~\cite{Gattringer:2002mr} is obtained 
for $N = 20$ modes of the overlap operator.

We take this agreement as a criterium for choosing the number of modes $N$ 
and we use further exclusively $N = 20$ modes 
of the overlap operator in calculations of correlation functions.
The three-dimensional
dyon cluster density in this case is equal to $\rho=3.03$~fm$^{-3}$. 
Note, that respective dimensionless 
value $\rho/T^3=0.98$ should be compared with the value 0.74 obtained 
in Ref.~\cite{Larsen:2015vaa} for the density of dyons in case of the
SU(2) dyon model at a temperature close to $T_c$.

\begin{figure*}[!htb]
\centering
\hspace{0.1cm} \includegraphics[width=.8\textwidth]{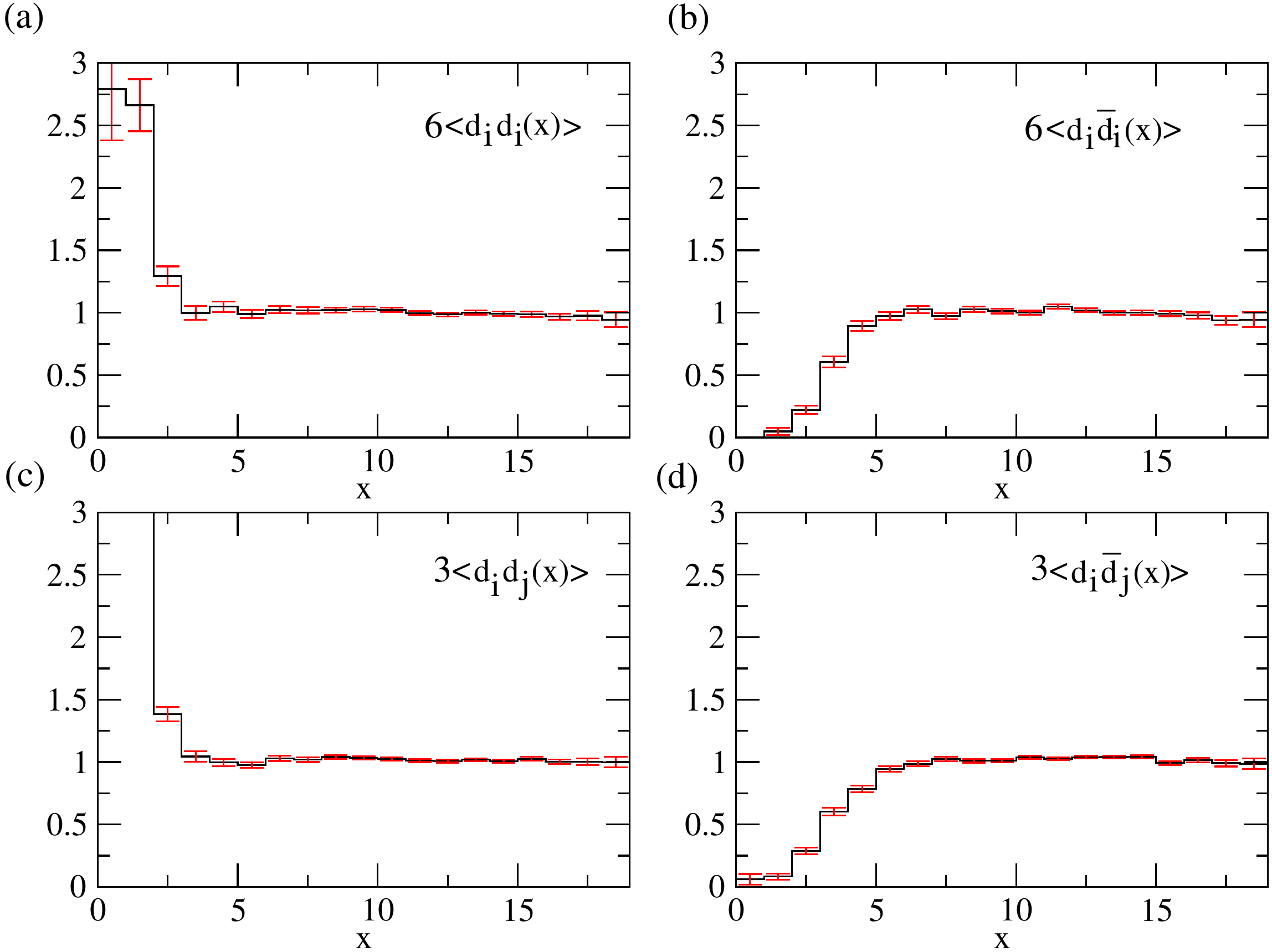}\\
\caption{Correlators (normalized to the total density of dyons and antidyons 
of all types squared) of dyon and antidyon densities are shown as histograms 
over distance with bins of one lattice spacing. The errors are shown in the 
centers of corresponding bins. }
\vspace{1cm}
\label{fig:correlations}
\end{figure*}

For each configuration 
we calculated the distances between topological lump objects of different 
types (dyons/antidyons of three types):
two dyons (or antidyons) of the same type ($d_i~d_i$),
dyon and antidyon of the same type ($d_i~{\bar d}_i$), 
two dyons (or antidyons) of different type ($d_i~d_j$), 
dyon and antidyon of different type ($d_i~{\bar d}_j$).
The numbers of these pairs in bins of (3-dimensional) distances 
from $x$ to $x+dx$ divided by the numbers of lattice points falling 
in the same bins are presented as functions of the calculated distances 
over the range from zero distance to the maximal distance 
$\sqrt{(24/2)^2+(24/2)^2+(24/2)^2}\approx 20$ lattice units. 
Normalized (to the total density of dyons and antidyons of all types 
squared) the correlators of dyons and antidyons densities  
\beq
<d_id_i(x)>\equiv\frac{\sum_i <d_i(x)d_i(0)+\bar{d}_i(x)\bar{d}_i(0)>}{(\sum_i <d_i+\bar{d}_i>)^2}
\label{eq:corr_a}
\eeq
\beq
<d_i \bar{d}_i(x)>\equiv\frac{\sum_i <d_i(x)\bar{d}_i(0)+\bar{d}_i(x){d}_i(0)>}{(\sum_i <d_i+\bar{d}_i>)^2}
\label{eq:corr_b}
\eeq
\beq
<d_id_j(x)>\equiv\frac{\sum_{i\ne j} <d_i(x)d_j(0)+\bar{d}_i(x)\bar{d}_j(0)>}{(\sum_i <d_i+\bar{d}_i>)^2}
\label{eq:corr_c}
\eeq
\beq
<d_i\bar{d}_j(x)>\equiv\frac{\sum_{i\ne j} <d_i(x)\bar{d}_j(0)+\bar{d}_i(x){d}_j(0)>}{(\sum_i <d_i+\bar{d}_i>)^2}
\label{eq:corr_d}
\eeq
are shown in Fig.~\ref{fig:correlations} as histograms with bins of  
one lattice spacing size. The errors are shown in the centers of the 
corresponding bins.

In Fig.~\ref{fig:correlations} the first two bins are not shown. 
Respective correlator values are equal to
51.6 and 4.47  for first and second bins, correspondingly.
We see that two dyons (or antidyons) of different type are attracting 
(positively correlated) at small distances. 
At this point we are in agreement
with model results of Ref.\cite{Larsen:2015vaa}. As a result
of this attraction
the half of dyons and antidyons are combined in dyon pairs and dyon  
triplets (full calorons).
We see also some attraction at small distance for two dyons (or antidyons)
of the same type 
(which cannot form dyon pairs and full calorons)
(see Fig.~\ref{fig:correlations}a) 
This is in contrast to the repulsion postulated in Ref.\cite{Larsen:2015vaa}.
As for dyon and antidyon interaction we observe a repulsion at small distances 
of strength independent of the types of dyons and antidyons 
(see Fig.~\ref{fig:correlations}b and Fig.~\ref{fig:correlations}d). 
At larger distances in all four cases we do not see any interaction
(non-trivial correlation). 

We note that our results are qualitatively the same for $N=10$ and 30.

\section{Conclusions}
\label{sec:conclusions}

In $SU(3)$ lattice gauge theory, using a small number of modes of the overlap 
Dirac operator with eigenvalues closest to zero, we have investigated clusters 
formed by the 
UV-filtered fermionic topological charge density. 
The topological charge density has been computed for 
three different types of temporal boundary conditions 
applied to the overlap Dirac operator.
Assuming that these clusters correspond to dyons, we have obtained
their frequency of occurrence and demonstrated the tendency to combine 
into triplets (calorons) or to form pairs of dyons apart from remaining 
isolated dyons.
 
We accomplished a first lattice computation of the dyon correlation 
functions defined in eqs.~(\ref{eq:corr_a}-\ref{eq:corr_d}).
We found at small distances attraction for two dyons (or antidyons) 
and repulsion for dyon and antidyon.
The attraction for two dyons (or antidyons) of different type is larger 
than the attraction for two dyons (or antidyons) of same type. 
Repulsion for dyon and antidyon does not depend on types of dyons and 
antidyons.
At larger distances in all cases we do not see any correlations.

\subsection*{Acknowledgments}
V.G.B. and B.V.M. are supported by the RFBR grant 18-02-40130.


\bibliographystyle{apsrev}


\end{document}